\documentclass[final,5p,times,twocolumn]{elsarticle}
\usepackage{amssymb,natbib,graphics}
\def\kB{k_{\rm B}}
\def\sr{\skew3\dot\gamma}
\def\bbbr{{\rm I\!R}}
\def\bbbz{{\sf Z\hspace{-0.4 em}Z}}
\def\bbbzi{{\sf Z\hspace{-0.3 em}Z}}

\journal{Journal of Non-Newtonian Fluid Mechanics}

\begin{document}

\begin{frontmatter}

\title{Lie groups in nonequilibrium thermodynamics: Geometric structure behind viscoplasticity}

\author{Hans Christian \"Ottinger}

\address{ETH Z\"urich, Department of Materials, Polymer Physics,
HCI H 543, CH-8093 Z\"urich, Switzerland}

\ead{hco@mat.ethz.ch} 

\begin{abstract}
Poisson brackets provide the mathematical structure required to identify the reversible contribution to dynamic phenomena in nonequilibrium thermodynamics. This mathematical structure is deeply linked to Lie groups and their Lie algebras. From the characterization of all the Lie groups associated with a given Lie algebra as quotients of a universal covering group, we obtain a natural classification of rheological models based on the concept of discrete reference states and, in particular, we find a clear-cut and deep distinction between viscoplasticity and viscoelasticity. The abstract ideas are illustrated by a naive toy model of crystal viscoplasticity, but similar kinetic models are also used for modeling the viscoplastic behavior of glasses. We discuss some implications for coarse graining and statistical mechanics.
\end{abstract}


\end{frontmatter}

\section{Which mathematics behind which physics?}
The purpose of this work is to explore an elaborate mathematical
framework and its basic concepts in order to illuminate some
physics and to develop a natural and general scenario for a range
of phenomena. The underlying mathematics is the abstract theory of
Lie groups and, in particular, the correspondence between
\emph{Lie groups and Lie algebras}. Our concrete interest is in
groups of space transformations. The physics associated with space
transformations is the \emph{theory of deformation of materials},
ranging from fluids to solids and encompassing all kinds of
intermediate behavior. We are thus dealing with elasticity theory
and rheology, viscoelasticity, viscoplasticity, and all that. We
here make an attempt at classifying models and phenomena in
rheology from a mathematical perspective.
The key idea is that we identify discrete groups of
equivalent reference states that evolve slowly in time, much
slower than the structural variables describing deviations from
the reference states. We lay the foundations for the kinetic
theory of systems with reference states so that it becomes
possible to bridge the gap between atomistic and phenomenological
models of viscoplasticity.

Lie algebras are known to play an important role in the theory of
complex fluids \citep{hcobet}. Lie algebras allow us to introduce
Poisson brackets and Poisson brackets allow us to introduce
reversible dynamics (see Sec.~\ref{seceqBET} for details). With
the help of a Poisson bracket, an energy function generates
Hamiltonian dynamics, which is the prototype of mechanistically
controlled or reversible dynamics, to be separated from the
irreversible ``rest'' in any thermodynamics. \emph{Nonequilibrium
thermodynamics can take place only in structured spaces}, more
precisely, thermodynamic systems must possess a Poissonian
structure. This geometric structure is rooted in Lie algebras. More
carefully, one should consider not only Lie algebras but also Lie
groups, because one is usually working in infinite-dimensional
spaces: Not every infinite-dimensional representation of a Lie
algebra is guaranteed to exponentiate to a representation of
some group (see \S 6.5 of \citet{Rossmann}). The relevant Poisson brackets are
actually obtained by a procedure known as Lie-Poisson reduction
from the canonical Poisson bracket on the cotangent bundle of a
Lie group \citep{MarsdenRatiu,hcobet}, so that a group, or a
representation of a group, is the true starting point for
providing structure in nonequilibrium thermodynamics even though
the final Poisson bracket is given in terms of the Lie algebra.

Once one is aware of the fact that thermodynamics requires
structured spaces associated with Lie groups and their Lie
algebras, it is natural to look at all possible Lie groups
associated with a given Lie algebra. These groups can actually be
characterized very nicely \citep{Hamermesh,Rossmann}. There is a
unique simply connected universal covering group $\mathcal G$, and
all groups with the same Lie algebra are isomorphic to a quotient
group of the form ${\mathcal G}/{\mathcal Z}$, where $\mathcal Z$
is a finite or infinite discrete normal subgroup of $\mathcal G$
(a subgroup ${\mathcal Z} \subset {\mathcal G}$ is normal if $g z
g^{-1} \in {\mathcal Z}$ for all $z \in {\mathcal Z}$ and $g \in
{\mathcal G}$). Actually, $\mathcal Z$ must be contained in the
center of the group $\mathcal G$, that is, all elements of
$\mathcal Z$ commute with all elements of $\mathcal G$ (or, $g z
g^{-1} = z$; see \S 2.6 of \citet{Rossmann}). What is the physical
significance of the groups $\mathcal Z$ and ${\mathcal
G}/{\mathcal Z}$ in our application of group theory? Why can we
interpret the elements of the discrete group $\mathcal Z$ as
reference states?

For illustration, let us consider the subgroup of unidirectional
shear deformations of a simple cubic lattice in one of its
principal directions, ${\mathcal G} = \bbbr$, with addition as the
binary group operation. For small shear deformations, we expect elastic
behavior. For a unit shear deformation between neighboring lattice
planes, we reach a state that is fully equivalent to the
undeformed state and, therefore, such a unit shear deformation is
plastic. The discrete normal subgroup of integers, ${\mathcal Z} =
\bbbz$, represents a set of equivalent reference states, and
$\bbbr / \bbbz = I_0 = [ -1/2, 1/2[$ describes shear deformations
with respect to a reference state. The topology of the Lie group
$\bbbr / \bbbz$ (with the identification of $-1/2$ and $1/2$) is
that of a circle, which is not simply connected. The occurrence of
discrete subgroups is a general mathematical feature and it is
natural to associate them with the equivalent reference
configurations of a lattice. Only the simply connected universal
covering group is free of reference states and hence provides the
natural setting for viscoelastic liquids and other complex fluids.
All smaller groups are associated with reference states and
viscoplasticity.

The detailed elaboration of the above picture is the purpose of the following sections, after compiling some basic knowledge from group theory (Sec.~\ref{seceqgroup}) and nonequilibrium thermodynamics (Sec.~\ref{seceqBET}). By means of a toy example, we illustrate modeling on the full group (Sec.~\ref{sectoyfull}) and on the quotient group (Sec.~\ref{sectoyquot}) and observe that modeling on the quotient group is much simpler but requires some additional ingredients. A classification of rheological models resulting from the elaborated picture is offered in the final discussion (Sec.~\ref{secdiscussion}).

\section{Some basic equations for groups} \label{seceqgroup}
The Lie group of interest in describing the response of materials
to stress is the group of smooth space transformations. If we
consider a domain $D \subset \bbbr^3$, the group of space
transformations is given by
\begin{equation}\label{Gdef}
  {\mathcal G} = \{ g: D \rightarrow D \, | \, g \mbox{ smooth},
  g \mbox{ bijective} \} ,
\end{equation}
with the composition of functions as the binary group operation.
In the following sections, we mostly consider the subgroup of
unidirectional shear transformations on $D = \bbbr^3$,
\begin{equation}\label{Gdefshear}
  {\mathcal G}_{\rm s} = \Bigg\{ g_{\rm s}: \bbbr^3 \rightarrow \bbbr^3 ,
  \quad \left( \begin{array}{c}
  x \\ y \\ z \end{array} \right) \mapsto
  \left( \begin{array}{c}
  x + \Delta(y) \\ y \\ z \end{array} \right) \Bigg\} ,
\end{equation}
where the binary operation of composition leads to the addition of
the displacements $\Delta(y)$. The subgroup ${\mathcal G}_{\rm s}$
hence is commutative. The local shear strain is given by
$\gamma(y) = d \Delta(y)/d y$. For homogeneous shear deformations,
we have $\Delta = \gamma y$ with constant $\gamma$.

As small transformations can be visualized as effected by a velocity field acting over a short time, the Lie algebra associated with the group ${\mathcal G}$ of space transformations on $D$ provides the Poisson structure on the space of velocity fields on $D$ \citep{MarsdenRatiu,hcobet}. In the theory of complex fluids, the internal structure of the fluids is described by a vector space of functions defined on the same spatial domain as the velocity field,
\begin{equation}\label{Vdef}
  V = \{ v_{\rm struc}: D \rightarrow \bbbr^m \} .
\end{equation}
For the simplest case of Newtonian fluids, we have $m=2$ for the
number of structural variables because mass density and
temperature are sufficient to describe the local state of the
moving fluid. For complex fluids, additional fields are needed to
characterize the internal state of the fluid \citep{hcobet}. To
introduce structure on the space $V$ one needs an action of ${\mathcal G}$
on the dual of $V$, which is also a vector space. When Lie-Poisson
reduction is applied to the semidirect product of ${\mathcal G}$
and the dual of $V$, one obtains an extension of the Poisson
structure from the space of velocity fields to the \emph{space of
structural variables} $V$ \citep{MarsdenRatiu,hcobet}. Typically,
a space transformation $g \in {\mathcal G}$ acts on a real- or
vector-valued function on $D$ by transforming the arguments of the
function. For real valued functions, the values at the transformed
positions may be unchanged or multiplied by the Jacobian of the
transformation, depending on whether we deal with a scalar or
scalar density field. For vector or tensor functions, in addition,
the components at the transformed positions may be mixed by proper
tensor transformation laws.

By introducing velocity fields and the structural variables (\ref{Vdef}) in a region of space $D$, we have chosen to use an Eulerian description of complex fluids. For a deeper understanding of the required Poisson brackets we can start with a Lagrangian description of fluids, where the particle relabeling symmetry is the key to successful reduction to the smaller space of the Eulerian description (see \citep{Salmon88}, Section 1.5 of \citep{MarsdenRatiu}, Appendix B.4 of \citep{hcobet}, and references therein).

As mentioned before, the theory of complex fluids is based on functions defined on the same domain $D$ as the group of space transformations ${\mathcal G}$. These functions give the values of the structural variables at each point occupied by the fluid. Another natural possibility to construct properly structured spaces employs functions defined on the group  ${\mathcal G}$ itself, or some homomorphic group. More precisely, we consider the vector space of integrable real-valued functions (the generalization to vector-valued functions is straightforward),
\begin{equation}\label{Pdef}
  I = \{ f: {\mathcal G} \rightarrow \bbbr \} .
\end{equation}
The definition of $I$ assumes a concept of integration and hence a
measure on the Lie group $\mathcal G$. The natural choice is the
Haar measure defined on all locally compact topological groups,
which is invariant under the action of the group (see \S 5.2 of
\citet{Rossmann}). Up to a positive multiplicative constant, there
exists a unique left-invariant and right-invariant Haar measure
(in general, the two measures do not coincide). In particular, we
are interested in the subset $P \subset I$ of nonnegative
functions with integral unity which can be interpreted as the set
of probability densities for finding a particular space
transformation or state of deformation of a (piece of) material.
In kinetic theory, such probability densities are also referred to
as \emph{configurational distribution functions}. In a stochastic
interpretation, the evolution of the configurational distribution
function can equivalently be considered as a stochastic process in
the underlying Lie group $\mathcal G$ (in particular,
Fokker-Planck or diffusion equations for configurational
distribution functions are equivalent to stochastic differential
equations \citep{Gardiner,hcobook}). For local theories of the
deformation behavior of materials, it may be sufficient to
consider the subgroup of homogeneous space transformations (to be
used separately at each point in space).

Given a probability density $p \in P$ on $\mathcal G$, one can
naturally define a probability density $p_{\mathcal Z}$ on
the quotient group ${\mathcal G}/{\mathcal Z}$ by superimposing the contributions
from all equivalence classes,
\begin{equation}\label{probinduced}
  p_{\mathcal Z}(g) = \sum_{z \in {\mathcal Z}} p(zg) .
\end{equation}
However, in general, there is no ``periodic continuation'' of a
given probability density on ${\mathcal G}/{\mathcal Z}$ because
$\mathcal Z$ may be an infinite group so that there arises a
problem with normalization. Moreover, one would need suitable
``periodic boundary conditions'' to obtain a smooth continuation.
We denote the averages of a random variable $f$ on $\mathcal G$ or
${\mathcal G}/{\mathcal Z}$ with respect to the probability
density $p$ or $p_{\mathcal Z}$ as $\langle f \rangle$ or $\langle
f \rangle_{\mathcal Z}$, respectively. If a function $f$ possesses
the ``periodicity property'' $f(zg) = f(g)$ for all $z \in
{\mathcal Z}$ and $g \in {\mathcal G}$, then we have $\langle f
\rangle = \langle f \rangle_{\mathcal Z}$. We here use the
language of Fourier transforms because there exists a
well-developed theory of harmonic analysis on locally compact
groups, where the Fourier transform takes functions on a group to
functions on the dual group, with particularly powerful results
for commutative or compact groups. In the case of unidirectional
shear transformations of lattices, we deal with standard Fourier
analysis and periodic solutions.

\section{Background from nonequilibrium thermodynamics}  \label{seceqBET}
Time-evolution equations for nonequilibrium systems have a
well-defined structure in which reversible and irreversible
contributions are identified separately. As pointed out before,
the reversible contribution is generally assumed to be of the
Hamiltonian form and hence requires an underlying geometric
structure which reflects the idea that the reversible time
evolution should be ``under mechanistic control.'' The remaining
irreversible contribution is driven by the gradient of a
nonequilibrium entropy. We need a separate geometric structure to be obtained by entirely different arguments; however, for preserving symmetries under coarse graining, group representations play an important role also in the construction of the proper geometric structure for generating the irreversible contribution to time evolution (see Section 6.1.6 of \citep{hcobet}).

Our discussion is based on the GENERIC (``general equation for the
nonequilibrium reversible-irreversible coupling'') formulation of
time-evolution for nonequilibrium systems
\citep{hco99,hco100,hcobet},
\begin{equation} \label{LMformulation}
  \frac{dx}{dt} = L \frac{\delta E}{\delta x} +
  M \frac{\delta S}{\delta x} ,
\end{equation}
where $x$ represents the set of independent variables required for
a complete description of a given nonequilibrium system, $E$ and
$S$ are the total energy and entropy expressed in terms of the
variables $x$, and $L$ and $M$ are certain linear operators, or
matrices, which can also depend on $x$. The two contributions to
the time evolution of $x$ generated by the total energy $E$ and
the entropy $S$ in Eq.~(\ref{LMformulation}) are the reversible
and irreversible contributions, respectively. Because $x$
typically contains position-dependent fields, such as the local
mass, momentum and energy densities of hydrodynamics, the state
variables are usually labeled by continuous (position) labels in
addition to discrete ones. A matrix multiplication, which can
alternatively be considered as the application of a linear
operator, hence implies not only summations over discrete indices
but also integrations over continuous labels, and $\delta/\delta
x$ typically implies functional rather than partial derivatives.
Equation (\ref{LMformulation}) is supplemented by the
complementary degeneracy requirements
\begin{equation} \label{LSconsistency}
  L \frac{\delta S}{\delta x}=0
\end{equation}
and
\begin{equation} \label{MEconsistency}
  M \frac{\delta E}{\delta x}=0 .
\end{equation}
The requirement that the entropy gradient $\delta S/\delta x$ is
in the null-space of $L$ in Eq.~(\ref{LSconsistency}) expresses
the reversible nature of the $L$-contribution to the dynamics: the
functional form of the entropy is such that it cannot be affected
by the operator generating the reversible dynamics. The
requirement that the energy gradient $\delta E/\delta x$ is in the
null-space of $M$ in Eq.~(\ref{MEconsistency}) expresses the
conservation of the total energy in a closed system by the
$M$-contribution to the dynamics.

Further general properties of $L$ and $M$ are discussed most
conveniently in terms of the Poisson and dissipative brackets
\begin{equation} \label{Poissonbrackdef}
  \{A,B\} =  \frac{\delta A}{\delta x} L
  \frac{\delta B}{\delta x} ,
\end{equation}
\begin{equation} \label{dissipbrackdef}
  [A,B] = \frac{\delta A}{\delta x} M
  \frac{\delta B}{\delta x} ,
\end{equation}
where $A$, $B$ are sufficiently regular real-valued functions on
the space of independent variables. In terms of these brackets,
Eq.~(\ref{LMformulation}) and the chain rule lead to the following
time-evolution equation of an arbitrary function $A$ in terms of
the two separate generators $E$ and $S$,
\begin{equation} \label{brackform}
  \frac{dA}{dt} = \{A,E\} + [A,S] .
\end{equation}
The further conditions for $L$ can now be stated as the
antisymmetry property
\begin{equation} \label{condLasym}
  \{A,B\}=-\{B,A\} ,
\end{equation}
the product or Leibniz rule
\begin{equation} \label{Leibnizrule}
  \{A B,C\} = A \{B,C\} + B \{A,C\} ,
\end{equation}
and the Jacobi identity
\begin{equation} \label{condLJacobi}
  \{A,\{B,C\}\}+\{B,\{C,A\}\}+\{C,\{A,B\}\}=0 ,
\end{equation}
where $C$ is another arbitrary sufficiently regular real-valued
function on the state space. The Jacobi identity
(\ref{condLJacobi}), which is a highly restrictive condition for
formulating proper reversible dynamics, expresses the invariance
of Poisson brackets in the course of time (time-structure
invariance). All these properties are well-known from the Poisson
brackets of classical mechanics, and they capture the essence of
reversible dynamics.

Further properties of $M$ can be formulated in terms of the
symmetry condition
\begin{equation} \label{condMsym}
  [A,B]=[B,A] ,
\end{equation}
and the non-negativeness condition
\begin{equation} \label{condMpos}
  [A,A] \ge 0 .
\end{equation}
This non-negativeness condition, together with the degeneracy
requirement (\ref{LSconsistency}), guarantees that the entropy is
a nondecreasing function of time,
\begin{equation} \label{increntrop}
  \frac{dS}{dt} = \frac{\delta S}{\delta x} M
  \frac{\delta S}{\delta x} = [S,S] \ge 0 .
\end{equation}
The properties (\ref{condMsym}) and (\ref{condMpos}) imply the
symmetry and the positive-semidefiniteness of $M$ (for a more
sophisticated discussion of the Onsager-Casimir symmetry
properties of $M$, see Sections 3.2.1 and 7.2.4 of
\citet{hcobet}). From a physical point of view, $M$ may be
regarded as a friction matrix (actually, often rates or inverse
friction coefficients occur in $M$).

\section{Toy model on universal covering group} \label{sectoyfull}
We illustrate the general ideas of Lie algebras and quotient
groups in the context of thermodynamic modeling for a toy model of
crystal viscoplasticity \citep{hco176}. We consider the
unidirectional shear deformations (\ref{Gdefshear}) of a simple
cubic crystal in one of the principal directions caused by shear
stresses on the boundaries. The group ${\mathcal G}_{\rm s}$ of
Eq.~(\ref{Gdefshear}) is fully characterized by the unidirectional
displacements $\Delta(y)$. For homogeneous deformations, the group
is isomorphic with the additive group of real numbers, $\gamma =
[\Delta(y)/y] \in \bbbr$, and the set of equivalent reference
states is given by the discrete normal subgroup of integers
${\mathcal Z} = \bbbz$. In this section, we model on the simply
connected universal covering group ${\mathcal G} = \bbbr$ of all
shear deformations. This type of modeling on the full space is
familiar from the theory of complex fluids, avoids the formulation
of boundary conditions, and provides the most complete information
about the evolution of reference states and relative deformations.
Direct modeling on the quotient group $\bbbr/\bbbz$ is the topic
of the subsequent section.

Our toy model should not be confused with a realistic model of crystal viscoplasticity. It is well-known that the key to understanding crystal deformation are dislocations. However, the dynamics of dislocations involves the shearing of crystal domains and our toy model could hence offer a deeper understanding of the parameters (such as viscosity) in phenomenological theories of dislocation dynamics \cite{Hutter09}. Surprisingly, our model is also closely related to theories of shear transformation zones in metallic glass-forming liquids. For amorphous systems, periodic functions are used to model the \emph{average} potential energy as an idealization of a much more complicated energy landscape with many minima on a wide range of energy levels \cite{Johnsonetal07}.

\subsection{Thermodynamic formulation}
In order to formulate a toy model of crystal viscoplasticity, we
introduce the momentum density field, which has a nonzero
component only in the $x$-direction and depends only on the
$y$-coordinate, $g(y)$, the internal energy density $\epsilon(y)$,
and the probability density $p(\gamma,y)$ over the shear
deformations $\gamma$ between adjacent layers of a simple cubic
lattice with spacing $a$. For every $y$, the configurational
distribution function $p(\gamma,y)$ is a probability density.

The total energy is given as the sum of kinetic, internal, and
configurational energy contributions,
\begin{equation}\label{toyE}
  E = N a^2 \int \left[ \frac{g(y)^2}{2 \rho} + \epsilon(y) \right] dy
  + \frac{1}{a} \int \int p(\gamma,y) \Phi(\gamma) d\gamma dy ,
\end{equation}
where $N$ is the number of atoms in a single displaced layer, $N
a^2$ is the area of this layer, and $\Phi$ is the potential energy
associated with shearing a whole layer. For shear flows, a
constant initial density $\rho$ does not change with time and
hence plays the role of a fixed parameter. The idea of equivalent
reference states requires a periodic potential energy function
$\Phi$ with period unity. In other words, the potential energy
$\Phi$ is invariant under shear deformations $\gamma \in \bbbz$.
The prototypical example of such a potential $\Phi$ is
\begin{equation}\label{potential}
  \Phi_\varepsilon(\gamma) = \frac{H}{2\pi^2} \,
  \frac{1}{(\cos \pi \gamma)^2 + \varepsilon} ,
\end{equation}
with the force
\begin{equation}\label{force}
  F_\varepsilon(\gamma) = - \frac{H}{2\pi} \,
  \frac{\sin 2\pi \gamma}{[(\cos \pi \gamma)^2 + \varepsilon]^2} .
\end{equation}
The parameter $H$ has the dimensions of energy; $H \gamma^2 / 2$
is the total energy for small shear displacements $\gamma$ between
two neighboring layers and small $\varepsilon$. The parameter $H$
hence has an extensive character, $H= \hat{H} N$, where $\hat{H}$
characterizes the energy per atom. The shear modulus is given by
$G = \hat{H}/a^3$ so that, for known lattice spacing $a$,
$\hat{H}$ can be obtained from a mechanical measurement in the
elastic regime. The larger the number of atoms in the layer, the
larger is the energy required to achieve shear deformations, and
the smaller are the thermal shear fluctuations for a given value
of $\kB T$. The parameter $\varepsilon$ allows us to make the
energy barrier between the different minima for each atom very
high. In particular, the barrier becomes insurmountable for
$\varepsilon=0$. The control parameter $\varepsilon$ hence is the
key to identifying well-defined, long-living reference states and
deformations with respect to those.

The total entropy is given as the sum of a thermal and a
configurational contribution,
\begin{equation}\label{toyS}
  S = N a^2 \int s(\epsilon(y)) dy - \frac{1}{a} \kB
  \int \int p(\gamma,y) \ln p(\gamma,y) d\gamma dy ,
\end{equation}
where $s(\epsilon)$ is the density of the entropy associated with
internal energy and $\kB$ is Boltzmann's constant. The
configurational entropy contribution has the well-known Boltzmann
form.

Before we can evaluate the functional derivatives of energy and
entropy, we need to define the relevant linear spaces more
carefully. In the spirit of Lie-Poisson reduction, we consider the
functional derivatives of observables as elements of an underlying
Lie algebra, and variations of the independent fields as elements
of the dual Lie algebra. For every $y$, the normalization of the
configurational distribution function $p(\gamma,y)$ implies the
following constraint for elements of the dual of the Lie algebra:
\begin{equation}\label{constraintnorm}
  \int \delta p(\gamma,y) d\gamma = 0 .
\end{equation}
There is a further, more subtle constraint we need to impose on
$\delta p(\gamma,y)$. The configurational distribution function
describes the probability for finding a shear deformation between
any two layers, without considering any coupling between different
pairs of layers. However, the total imposed shear strain has to be
distributed over all the layers of the deformed crystal. We hence
impose the further constraint that there is no freedom of varying
the total strain,
\begin{equation}\label{constraintstrain}
  \int \int \gamma \, \delta p(\gamma,y) d\gamma dy = 0 .
\end{equation}
We next need to define the constraints on the Lie algebra which
has the physical interpretation of the space of observables (the
pairing of the Lie algebra and its dual corresponds to the
averaging of observables). In defining the functional derivative
of some functional $A$ of the configurational distribution
function, the constraints (\ref{constraintnorm}) and
(\ref{constraintstrain}) imply the following freedom:
\begin{equation}\label{cfreedom}
  \frac{\delta A}{\delta p(\gamma,y)} \rightarrow
  \frac{\delta A}{\delta p(\gamma,y)} + f(y) + C \gamma ,
\end{equation}
with an arbitrary function $f(y)$ and a constant $C$, which play
the role of Lagrange multipliers which still need too be fixed. In the spirit of reciprocal base
vectors, where the dual of $1$ is $1$ and the dual of $\gamma$ is
the differential operator $\partial / \partial \gamma$, we use
this freedom to satisfy the dual conditions
\begin{equation}\label{LAcons1}
  \int p(\gamma,y) \frac{\delta A}{\delta p(\gamma,y)} d\gamma = 0 ,
\end{equation}
and
\begin{equation}\label{LAcons2}
  \int \int p(\gamma,y) \frac{\partial}{\partial\gamma}
  \frac{\delta A}{\delta p(\gamma,y)} d\gamma dy = 0 .
\end{equation}
Equations (\ref{LAcons1}) and (\ref{LAcons2}) are the constraints
on the elements of the underlying Lie algebra (of functional
derivatives), whereas (\ref{constraintnorm}) and
(\ref{constraintstrain}) are the corresponding constraints on the
elements of the dual of the Lie algebra (of variations of the
independent fields). The constraint (\ref{LAcons2}) plays a
crucial role in obtaining a meaningful evolution equation for the
configurational distribution function because it makes sure that
the imposed macroscopic shear deformation is accommodated by the
crystal. Motivated by the form of these constraints, we introduce
two kinds of averages: $\langle \cdots \rangle$ for an average
performed with the configurational distribution function $p$, and
$\langle \cdots \rangle_{\rm sp}$ for spatial averaging in the $y$
direction.

We can now evaluate the suitably normalized functional derivatives
of energy and entropy (to be consistent with a three-dimensional
description of the system one needs to introduce appropriate
factors of area, $N a^2$, into our formulation for unidirectional
flows):
\begin{equation}\label{gradientE}
  \left( \begin{array}{c}
  \frac{\delta E}{\delta g(y)}\\
  \frac{\delta E}{\delta \epsilon(y)}\\
  \frac{\delta E}{\delta p(\gamma,y)}\end{array}
  \right) = \left( \begin{array}{c}
  v(y) \\ 1 \\ (Na^3)^{-1} \left[ \Phi(\gamma)
  - \langle \Phi \rangle
  - ( \gamma - \langle \gamma \rangle ) \left\langle \! \left\langle
  \frac{d\Phi}{d\gamma} \right\rangle \! \right\rangle_{\rm sp}
  \right] \end{array} \right) ,
\end{equation}
and
\begin{equation}\label{gradientS}
  \left( \begin{array}{c}
  \frac{\delta S}{\delta g(y)}\\
  \frac{\delta S}{\delta \epsilon(y)}\\
  \frac{\delta S}{\delta p(\gamma,y)}\end{array}
  \right) = \left( \begin{array}{c}
  0 \\ 1/T(y) \\ - (Na^3)^{-1} \kB [ \ln p(\gamma,y)
  - \langle \ln p \rangle ] \end{array} \right) ,
\end{equation}
where $v(y)$ is the $x$-component of the velocity field and $T(y)$
is the temperature field. Note that the last components in the
functional derivatives (\ref{gradientE}) and (\ref{gradientS})
satisfy the constraints (\ref{LAcons1}) and (\ref{LAcons2}).

From the general theory of Lie-Poisson reduction mentioned in the
introduction, we obtain the following Poisson bracket:
\begin{eqnarray}
  \{A,B\} &=& \int \int p(\gamma,y) \Bigg[
  \left( \frac{\partial}{\partial\gamma}
  \frac{\delta A}{\delta p(\gamma,y)} \right)
  \left( \frac{\partial}{\partial y}
  \frac{\delta B}{\delta g(y)} \right) \nonumber\\
  && \qquad - \left( \frac{\partial}{\partial\gamma}
  \frac{\delta B}{\delta p(\gamma,y)} \right)
  \left( \frac{\partial}{\partial y}
  \frac{\delta A}{\delta g(y)} \right) \Bigg]
  d\gamma dy .
\label{toyPoiB}
\end{eqnarray}
There are no convective contributions to this Poisson bracket
because the system properties do not change in the direction of
flow. The flow affects only the configurational distribution
function by changing the shear deformation $\gamma$. The total
entropy function is a degenerate function of this Poisson bracket
because the configurational entropy density depends on $\gamma$
only through $p$.

With the dissipative bracket, we express diffusion (or smoothing)
in the space of shear deformations,
\begin{eqnarray}
  [A,B] &=& \frac{1}{\kB Na^3} \int \int D \, p(\gamma,y)
  \nonumber\\ & \times &
  \left[ Na^3 \frac{\partial}{\partial\gamma}
  \frac{\delta A}{\delta p(\gamma,y)} -
  \frac{d\Phi(\gamma)}{d\gamma} \frac{\delta A}{\delta \epsilon(y)}
  + \left\langle \! \left\langle \frac{d\Phi}{d\gamma} \right\rangle
  \frac{\delta A}{\delta \epsilon} \right\rangle_{\rm sp}
  \right] \nonumber\\ & \times &
  \left[ Na^3 \frac{\partial}{\partial\gamma}
  \frac{\delta B}{\delta p(\gamma,y)} -
  \frac{d\Phi(\gamma)}{d\gamma} \frac{\delta B}{\delta \epsilon(y)}
  + \left\langle \! \left\langle \frac{d\Phi}{d\gamma} \right\rangle
  \frac{\delta B}{\delta \epsilon} \right\rangle_{\rm sp}
  \right] \nonumber\\ & \times & d\gamma dy ,
\label{toydissB}
\end{eqnarray}
where $D$ is an inverse time scale for the diffusion of entire
layers, which we expect to be proportional to $1/N$. The energy is
a degenerate functional of this positive-semidefinite, symmetric
bracket. At this point, we have formulated all the thermodynamic
building blocks of our toy model. We can now look at their
implications.

From the fundamental evolution equation (\ref{LMformulation}) of
the GENERIC framework \citep{hco99,hco100,hcobet}, we obtain the
following evolution equation for $p$:
\begin{equation}\label{diffeq}
  \frac{\partial p}{\partial t} =
  - \frac{\partial}{\partial \gamma} \left[ \left(
  \frac{\partial v}{\partial y} + D
  \left\langle \! \left\langle \frac{1}{\kB T}
  \frac{d \Phi}{d \gamma}
  \right\rangle \! \right\rangle_{\rm sp}
  - \frac{D}{\kB T}
  \frac{d \Phi}{d \gamma} \right) p \right]
  + \frac{\partial}{\partial \gamma} \left( D
  \frac{\partial p}{\partial \gamma} \right) .
\end{equation}
According to GENERIC, in the momentum balance equation, there
occurs the following elastic shear stress associated with the
potential $\Phi$,
\begin{equation}\label{shearstress}
  \sigma_{\rm e}(y) = \frac{1}{Na^3} \left\langle
  \frac{d\Phi}{d\gamma} \right\rangle .
\end{equation}

From now on, we consider the GENERIC toy model for isothermal
homogeneous shear flow, that is, the shear rate,
\begin{equation}\label{shearrate}
  \sr = \frac{\partial v}{\partial y} ,
\end{equation}
and the temperature $T$ are independent of the position $y$. As
also the shear stress (\ref{shearstress}) becomes independent of
$y$, the diffusion equation (\ref{diffeq}) for the time-dependent
configurational distribution function $p(\gamma)$ can be rewritten
as
\begin{equation}\label{diffeqhom}
  \frac{\partial p}{\partial t} =
  - \frac{\partial}{\partial \gamma} \left[ \left(
  \sr + \frac{\sigma_{\rm e}}{\hat{\eta}} - \frac{D}{\kB T}
  \frac{d \Phi}{d \gamma} \right) p \right]
  + \frac{\partial}{\partial \gamma} \left( D
  \frac{\partial p}{\partial \gamma} \right) ,
\end{equation}
with the viscosity parameter
\begin{equation}\label{formshstr}
  \hat{\eta} = \frac{\kB T}{a^3} \, \frac{1}{N D} .
\end{equation}
The anticipated dependence of the inverse time scale $D$ on $N$
mentioned after Eq.~(\ref{toydissB}) has been built into this
definition of $\hat{\eta}$; the parameter $\hat{\eta}$ can hence
be interpreted as the characteristic viscosity scale for
single-atom processes and should hence be of the order of the
viscosity of low-molecular-weight liquids. It is convenient to
summarize the drift terms in equation (\ref{diffeqhom}) by
introducing a potential with a constant external field,
\begin{equation}\label{potentialmodo}
  \tilde{\Phi}(\gamma) =
  \Phi(\gamma) - ( \hat{\eta} \sr + \sigma_{\rm e} ) Na^3 \gamma .
\end{equation}
The external field reflects the viscous and elastic stresses.

From the diffusion equation (\ref{diffeqhom}) for the
configurational distribution function, we obtain by averaging
\begin{equation}\label{identgamde}
  \frac{d}{dt} \langle \gamma \rangle = \sr .
\end{equation}
By introducing constraints we have indeed forced the crystal to
feel the full applied shear deformation. Given the integers
$\bbbz$ as a discrete normal subgroup of the additive group of
shear deformations leaving the potential energy invariant,
Eq.~(\ref{probinduced}) can be used to split the configurational
shear deformation $\langle \gamma \rangle$ into the elastic part
\begin{equation}\label{shearelastic}
  \langle \gamma \rangle_{\rm e} =
  \langle \gamma \rangle_\bbbzi =
  \int\limits_{-1/2}^{1/2} \gamma p_\bbbzi(\gamma) d\gamma =
  \sum_{z\in\bbbzi}
  \int\limits_{-1/2}^{1/2} \gamma p(z+\gamma) d\gamma ,
\end{equation}
and the plastic part
\begin{equation}\label{shearplastic}
  \langle \gamma \rangle_{\rm p} = \langle \gamma \rangle -
  \langle \gamma \rangle_\bbbzi = \sum_{z\in\bbbzi} z p_z
  \quad \mbox{with }
  p_z = \int\limits_{-1/2}^{1/2} p(z+\gamma) d\gamma .
\end{equation}
The elastic part feels only the shear deformation with respect to
a reference state; the plastic part is the shear associated with
the reference states themselves. From the diffusion equation
(\ref{diffeqhom}), we obtain
\begin{equation}\label{shearratepl}
  \frac{d}{dt} \langle \gamma \rangle_{\rm p} =
  \sr_{\rm p} = \left. \left[ \left( \sr
  + \frac{\sigma_{\rm e}}{\hat{\eta}} \right) p_\bbbzi
  -D \frac{\partial p_\bbbzi}{\partial \gamma} \right]
  \right|_{\gamma = 1/2} .
\end{equation}

In the following, we consider only the diffusion equation
(\ref{diffeqhom}) rather than the complete set of
thermodynamically admissible evolution equations. An important
consequence of the general geometric ideas for formulating
thermodynamically admissible models, however, is the fact that the
argument $\gamma$ of the probability density $p$ should be
recognized to represent the group of shear deformations $\bbbr$,
and that the configurational energy is invariant under shear
deformations from the discrete subgroup $\bbbz$.

\subsection{Solutions for $\varepsilon=0$}
The equilibrium solution (for $\sr = 0$ and $\sigma_{\rm e} = 0$)
of the diffusion equation (\ref{diffeqhom}) for the potential
$\Phi_\varepsilon$ is expected to be given by the Boltzmann
factors,
\begin{equation}\label{Boltzfacs}
  p^{\rm eq}(\gamma) \propto \exp \left\{
  - \frac{\Phi_\varepsilon(\gamma)}{\kB T} \right\} .
\end{equation}
For a periodic potential, however, such a solution is not
normalizable on the real axis $\bbbr$. We hence need a more
careful discussion and we start with the potential $\Phi_0$ for
$\varepsilon=0$, that is, with infinitely high energy barriers
between wells. Then, one should note that for all integers $z$
\begin{equation}\label{bc00}
  p(\gamma) \Big|_{\gamma = z + 1/2} = \left.
  \frac{\partial p(\gamma)}{\partial \gamma} \right|_{\gamma =
  z + 1/2} = 0 \qquad \mbox{for } z\in\bbbz ,
\end{equation}
so that the points $z + 1/2$ for $z\in\bbbz$ cannot be passed or
reached. We can hence look for separate solutions on each interval
\begin{equation}\label{intervalldef}
  I_z = \left[ z-\frac{1}{2} , z+\frac{1}{2} \right[
  \qquad \mbox{for } z\in\bbbz
\end{equation}
around any integer $z$. On any interval $I_z$, the solution
(\ref{Boltzfacs}) can be normalized,
\begin{equation}\label{normBoltzfacs1}
  p^{\rm eq}(\gamma) = \frac{1}{Z} \exp \left\{
  - \frac{\Phi_0(\gamma)}{\kB T} \right\} , \quad
  Z = \int\limits_{-1/2}^{1/2} \exp \left\{
  - \frac{\Phi_0(\gamma)}{\kB T} \right\} d\gamma ,
\end{equation}
where the normalization constant $Z$ is independent of $z$. The
most general equilibrium solution on the entire real axis $\bbbr$
is given by
\begin{equation}\label{normBoltzfacs2}
  p^{\rm eq}(\gamma) = p_z \frac{1}{Z} \exp \left\{
  - \frac{\Phi_0(\gamma)}{\kB T} \right\} , \quad
  \mbox{for } \gamma \in I_z  \mbox{ with }
  \sum_{z\in\bbbzi} p_z = 1 ,
\end{equation}
where the probability $p_z$ for a deformation in the interval
$I_z$ does not change in time and is hence given by the initial
conditions from which an equilibrium state is reached. For
nonpathological initial conditions, also time-dependent solutions
fulfilling Eq.~(\ref{bc00}) can be constructed independently on
each interval $I_z$ and pieced together.

We now apply a homogeneous shear flow that leads to a modified
force law (\ref{potentialmodo}) with a constant external force.
For the particular potential $\Phi_\varepsilon$ given in
Eq.~(\ref{potential}), we obtain
\begin{eqnarray}
  \tilde{\Phi}_\varepsilon(\gamma) &=& \Phi_\varepsilon(\gamma)
  - ( \hat{\eta} \sr + \sigma_{\rm e} ) Na^3 \gamma
  \nonumber\\ &=&
  Na^3 \left[ \frac{G}{2\pi^2} \,
  \frac{1}{(\cos \pi \gamma)^2 + \varepsilon}
  - ( \hat{\eta} \sr + \sigma_{\rm e} ) \gamma \right] .
\label{potentialmod}
\end{eqnarray}
For $\varepsilon=0$, we can repeat the previous construction of
steady state solutions fulfilling the consistency conditions
(\ref{bc00}) for any finite applied shear rate $\sr$. The solution
(\ref{normBoltzfacs1}) on the interval $I_z$ is replaced by
\begin{eqnarray}
  \tilde{p}^{\rm eq}(\gamma) &=& \frac{1}{Z_z} \exp \left\{
  - \frac{\tilde{\Phi}_0(\gamma)}{\kB T} \right\} , \nonumber\\
  Z_z &=& \exp \left\{ \frac{z}{D} \left( \sr
  + \frac{\sigma_{\rm e}}{\hat{\eta}} \right) \right\}
  \int\limits_{-1/2}^{1/2} \exp \left\{
  - \frac{\tilde{\Phi}_0(\gamma)}{\kB T} \right\} d\gamma ,
\label{normBoltzfacs1mod}
\end{eqnarray}
where the normalization constant $Z_z$ now depends on $z$. The
distribution on each interval $I_z$ is deformed by the flow, but
there are no transitions between different potential wells.
According to Eq.~(\ref{shearplastic}), we hence have $\sr_{\rm p}
= 0$. Under steady-state conditions, we have also $\sr_{\rm e} =
0$.

\subsection{Solutions for small $\varepsilon$}
If we proceed from $\varepsilon=0$ to small values of
$\varepsilon$, the barriers at $z+1/2$ become large but finite so
that there arises the possibility of transitions between
neighboring intervals. The transition rates are small so that the
probabilities $p_z$ change slowly. The evolution in each interval
is much faster so that a steady state is reached in each interval
before there is a substantial probability for transitions. To the
next level of approximation, the solutions of the diffusion
equation (\ref{diffeqhom}) are hence still given by
Eq.~(\ref{normBoltzfacs2}), but the probabilities $p_z$ are to be
obtained from a set of evolution equations in terms of the (small)
transition rates.

Let us consider transitions between the intervals $I_z$ and
$I_{z+1}$ in the absence of flow (for $\sr = 0$ and $\sigma_{\rm
e} = 0$), when the probability $p_z$ is different from $p_{z+1}$.
In view of this difference, the almost stationary solutions around
the peak at $z+1/2$ cannot be of the flux-free symmetric Boltzmann
form (\ref{Boltzfacs}). We need a skewed stationary solution that
can interpolate between two different levels of probability. The
solution to this problem is given by the following generalization
of the Boltzmann distribution in Eq.~(\ref{Boltzfacs}), which is
fundamental for the subsequent analysis:
\begin{equation}\label{fluxsols}
  p^{\rm st}(\gamma) = \exp \left\{
  - \frac{\Phi_\varepsilon(\gamma)}{\kB T} \right\}
  \left[ c - c' \int_{\gamma_0}^\gamma \exp \left\{
  \frac{\Phi_\varepsilon(\gamma')}{\kB T} \right\} d\gamma' \right] ,
\end{equation}
where $c$ and $c'$ are suitable constants fixing the normalization
and the flux. Note that $c$ depends on the choice of the lower
limit of integration, $\gamma_0$, whereas $c'$ does not; for small
$c'$, however, this dependence is weak. For the solution
(\ref{fluxsols}), the probability flux is given by the constant
\begin{equation}\label{probflux}
  -D \left( \frac{1}{\kB T}
  \frac{d \Phi_\varepsilon}{d \gamma} p^{\rm st}
  + \frac{\partial p^{\rm st}}{\partial \gamma} \right) = c' D ,
\end{equation}
so that we obtain a stationary solution of the diffusion equation
(\ref{diffeqhom}) with a constant probability flux through the
system. By using Eq.~(\ref{fluxsols}) around $z+1/2$, we obtain
for small differences between $p_z$ and $p_{z+1}$
\begin{eqnarray}
  \frac{p_z - p_{z+1}}{p_z} &=&
  \frac{p^{\rm st}(z) - p^{\rm st}(z+1)}{p^{\rm st}(z)} =
  \frac{c'}{c} \int_{(1/2)-\delta}^{(1/2)+\delta} \exp \left\{
  \frac{\Phi_\varepsilon(\gamma)}{\kB T} \right\} d\gamma
  \nonumber\\ &=& \frac{c'}{c} I =
  \frac{c'}{c} \varepsilon \sqrt{\frac{2\pi \kB T}{H}}
  \exp \left\{ \frac{H}{2 \pi^2 \kB T \varepsilon} \right\} ,
\label{cequation}
\end{eqnarray}
where the integral
\begin{equation}\label{Idef}
  I = \int_{(1/2)-\delta}^{(1/2)+\delta} \exp \left\{
  \frac{\Phi_\varepsilon(\gamma)}{\kB T} \right\} d\gamma ,
\end{equation}
when evaluated with the saddle point approximation, is independent
of the precise value of the small parameter $\delta$ (with
$\varepsilon \ll \delta \ll 1$) because it is dominated by the
high potential barrier around $\gamma = 1/2$. Equation
(\ref{cequation}) determines the small flux or flow rate parameter
$c'$. With this value of $c'$ and $c = p_z/Z$, we obtain the flux
of probability from $I_z$ to $I_{z+1}$. Taking into account the
change of $p_z$ also from the flux of probability from $I_{z-1}$
to $I_z$, the equations governing the slow time evolution of $p_z$
can then be written as
\begin{equation}\label{pzevol}
  \frac{d p_z}{d t} = r (p_{z+1} - 2 p_z + p_{z-1}) ,
\end{equation}
where the rate parameter $r$ is given by
\begin{equation}\label{rdef}
  r = \frac{D}{Z I} .
\end{equation}

In the presence of shear flow, matching solutions on neighboring
intervals can be constructed in the same way. A comprehensive
solution summarizing all the features to the discussed level of
precision can be written as
\begin{eqnarray}
  \renewcommand{\arraystretch}{2.5}
  && \tilde{p}(z+\gamma) = \frac{1}{Z_0} \exp \left\{
  - \frac{\tilde{\Phi}_\varepsilon(\gamma)}{\kB T} \right\}
  \nonumber\\ && \!\!\!\!\!\!\!\!\!\!\!\! \times
  \left\{ \begin{array}{l}
  p_z + \left( \frac{1}{w} p_{z+1} - w p_z \right) \frac{1}{I}
  \int_0^\gamma \exp \left\{
  \frac{\tilde{\Phi}_\varepsilon(\gamma')}{\kB T} \right\} d\gamma'
  \mbox{ for } \gamma \in [0, 1/2]\\
  p_z - \left( \frac{1}{w} p_z - w p_{z-1} \right) \frac{1}{I}
  \int^0_\gamma \exp \left\{
  \frac{\tilde{\Phi}_\varepsilon(\gamma')}{\kB T} \right\} d\gamma'
  \mbox{ for } \gamma \in [-1/2,0] \end{array}
  \right . ,
  \renewcommand{\arraystretch}{1}
\label{compactsolution}
\end{eqnarray}
on the interval $I_z$ (note that the solution on $I_z$ is
conveniently shifted to the interval $I_0$). The definition of the
integral $I$ of Eq.~(\ref{Idef}) remains unchanged and the
parameter $w$ is given by
\begin{equation}\label{wdefmod}
  w = \exp \left\{ \frac{1}{2 D} \left( \sr
  + \frac{\sigma_{\rm e}}{\hat{\eta}} \right) \right\} .
\end{equation}
The integral terms in Eq.~(\ref{compactsolution}) are negligible
except in narrow boundary layers. The values of the solution
(\ref{compactsolution}) on the boundaries are given by
\begin{equation}\label{limits}
  \renewcommand{\arraystretch}{2}
  \frac{1}{2 Z_0} \exp \left\{
  - \frac{\Phi_\varepsilon(1/2)}{\kB T} \right\} \times
  \left\{ \begin{array}{l}
  \frac{1}{w} p_{z+1} + w p_z
  \qquad \mbox{for } \gamma = 1/2 \\
  \frac{1}{w} p_z + w p_{z-1}
  \qquad \mbox{for } \gamma = - 1/2 \end{array}
  \right . ,
  \renewcommand{\arraystretch}{1}
\end{equation}
so that the continuous matching of the solution on neighboring
intervals becomes obvious. Moreover, we obtain
\begin{equation}\label{explprobmax}
  \left. \left[ \left( \sr
  + \frac{\sigma_{\rm e}}{\hat{\eta}} \right) p_\bbbzi
  -D \frac{\partial p_\bbbzi}{\partial \gamma} \right]
  \right|_{\gamma = 1/2} = r_0 \left( w - \frac{1}{w} \right) ,
\end{equation}
with the modified rate parameter
\begin{equation}\label{rdefmod}
  r_0 = \frac{D}{Z_0 I} .
\end{equation}

The modified time-evolution equations for $p_z$ in the presence of
flow become
\begin{equation}\label{pzevolmod}
  \frac{d p_z}{d t} = r_0 \left[ \frac{1}{w} p_{z+1} -
  \left( w + \frac{1}{w} \right) p_z + w p_{z-1} \right] .
\end{equation}
From Eqs.~(\ref{shearplastic}) and (\ref{pzevolmod}), we obtain
\begin{eqnarray}
  \sr_{\rm p} = r_0 \left( w - \frac{1}{w} \right) &=&
  2 r_0 \sinh \left[ \frac{1}{2 D} \left( \sr
  + \frac{\sigma_{\rm e}}{\hat{\eta}} \right) \right]
  \nonumber\\ &=&
  2 r_0 \sinh \left[ \frac{N a^3}{2 \kB T}
  ( \hat{\eta} \sr + \sigma_{\rm e} ) \right] ,
\label{drift}
\end{eqnarray}
which is consistent with Eqs.~(\ref{shearratepl}) and
(\ref{explprobmax}). Equation (\ref{drift}) relates the plastic
strain rate to the externally applied stress $\sigma_{\rm e}$
(note that $\hat{\eta} \sr$ is expected to be small compared to
$\sigma_{\rm e}$). It is closely related to Eyring's viscosity
formula \citep{Eyring36,ReeEyring55}. The ``volume of a flow
unit'' or ``viscosity volume'' is given by $N a^3$, where such a
large volume is a hallmark of highly viscous, plastic flow.
Accordingly, the rate parameter $r_0$ is extremely small. Equation
(\ref{drift}) is the main result of the asymptotic solution of our
toy model on the full universal covering group. Equation
(\ref{explprobmax}) shows that the configurational probability
density on the quotient group occurs naturally. The evolution
equation (\ref{pzevolmod}) implies the following H-theorem-like
identity for transitions between the different intervals,
\begin{eqnarray}
  \frac{d}{d t} \left( \sum_{z\in\bbbzi} p_z \ln \frac{p_z}{w^2}
  \right) &=& - r_0 \sum_{z\in\bbbzi}
  \left( \frac{p_{z+1}}{w} - w p_z \right) \nonumber\\
  & \times &
  \left[ \ln \frac{p_{z+1}}{w} - \ln (w p_z) \right] \leq 0 .
\label{entropyevol}
\end{eqnarray}

For a weak dependence of $p_z$ on $z$, Eq.~(\ref{pzevolmod}) can
be rewritten as a partial differential equation,
\begin{equation}\label{pzevolmodc}
  \frac{\partial p_z}{\partial t} = - r_0 \left( w - \frac{1}{w} \right)
  \frac{\partial p_z}{\partial z} + \frac{1}{2} r_0
  \left( w + \frac{1}{w} \right) \frac{\partial^2 p_z}{\partial z^2} .
\end{equation}
We once more realize that, on top of slow diffusion, there is a
small systematic flux of probability with the drift velocity
$\sr_{\rm p}$ given in Eq.~(\ref{drift}).

\section{Toy model on quotient group} \label{sectoyquot}
So far, we have developed a kinetic theory on the full, simply
connected group of shear deformations. Equation (\ref{pzevolmod})
describes the irreversible transitions between reference states.
In particular, it implies the plastic deformation rate
(\ref{drift}). We now address the construction of a complete
kinetic theory for the relative deformations, that is, on the
quotient group $I_0 = \bbbr/\bbbz$. We look for an approach that
reproduces the results of the preceding section, but works
directly and exclusively with a configurational distribution
function on the interval $I_0$.

\subsection{Key equations} \label{secke}
The general definition (\ref{probinduced}) of a probability
density on $I_0 = \bbbr/\bbbz$ for the group of shear deformations
reads
\begin{equation}\label{foldedsol}
  p_\bbbzi(\gamma) = \sum_{z\in\bbbzi} p(z+\gamma) \qquad
  \mbox{for } \gamma \in I_0 ,
\end{equation}
as already introduced in Eq.~(\ref{shearelastic}). Because of the
periodicity of the potential $\Phi$ and the linearity of the
diffusion equation (\ref{diffeqhom}) in $p$ for given $\sigma_{\rm
e}$, from any solution $p$ on the entire real axis $\bbbr$, one
obtains a solution $p_\bbbzi$ of the diffusion equation on $I_0$.
If the original solution is smooth, this construction ensures that
we obtain the same smoothness for the periodic boundary conditions
because one has identical values at $\gamma = - 1/2$ and $\gamma =
1/2$ in Eq.~(\ref{foldedsol}), as can be seen by shifting the
summation index $z$. A kinetic theory on $I_0$ must hence be based
on smooth periodic solutions of the diffusion equation, so that we
actually deal with a circle. Note that also the elastic stress
$\sigma_{\rm e}$ given in Eq.~(\ref{shearstress}) can be evaluated
as an average performed with $p_\bbbzi$,
\begin{equation}\label{shearstressZ}
  \sigma_{\rm e} = \frac{1}{Na^3} \left\langle
  \frac{d\Phi}{d\gamma} \right\rangle_\bbbzi ,
\end{equation}
because the averaged force is a periodic function of the shear
deformation.

In summary, we need to solve the following diffusion equation with
periodic boundary conditions on $I_0$:
\begin{equation}\label{diffeqhomZ}
  \frac{\partial p_\bbbzi}{\partial t} =
  - \frac{\partial}{\partial \gamma} \left[ \left(
  \sr + \frac{\sigma_{\rm e}}{\hat{\eta}} - \frac{D}{\kB T}
  \frac{d \Phi}{d \gamma} \right) p_\bbbzi \right]
  + \frac{\partial}{\partial \gamma} \left( D
  \frac{\partial p_\bbbzi}{\partial \gamma} \right) .
\end{equation}
From the diffusion equation (\ref{diffeqhomZ}), we obtain by
averaging
\begin{equation}\label{identgamdeZ}
  \frac{d}{dt} \langle \gamma \rangle_\bbbzi = \sr
  - \left. \left[ \left( \sr
  + \frac{\sigma_{\rm e}}{\hat{\eta}} \right) p_\bbbzi
  -D \frac{\partial p_\bbbzi}{\partial \gamma} \right]
  \right|_{\gamma = 1/2} = \sr - \sr_{\rm p} .
\end{equation}
We thus naturally recover the splitting of the total shear rate
into an elastic contribution and the plastic contribution
(\ref{shearratepl}). The model with a configurational distribution
function $p$ on the real line hence justifies the following
assumptions of the kinetic theory with $p_\bbbzi$ on the interval
$I_0$: (i) the use of smooth periodic boundary conditions for
$p_\bbbzi$, (ii) the shear stress $\sigma_{\rm e}$ given by
Eq.~(\ref{shearstressZ}), and (iii) the splitting of the shear
rate into elastic and plastic contributions, where the latter is
given in Eq.~(\ref{shearratepl}). All these assumptions have been
made in the original formulation of a toy model of crystal
viscoplasticity \citep{hco176}. Formally, the periodic theory
works perfectly well for time-dependent problems, even when the
energy barrier between wells is not assumed to be very large.

\subsection{Steady state solution}
For steady state situations, the solution of the diffusion
equation (\ref{diffeqhomZ}) is of the general form given in
Eq.~(\ref{fluxsols}). The constants have to be chosen such that we
obtain a periodic normalized solution. In particular, we can then
determine the plastic shear rate from Eq.~(\ref{identgamdeZ})
\citep{hco176}:
\begin{eqnarray}
  \sr_{\rm p} &=& \Bigg[ \int\limits_{-1/2}^{1/2}
  \exp \left\{ - \frac{\tilde{\Phi}(\gamma)}{\kB T} \right\}
  \Bigg( w \int\limits_\gamma^{1/2} \exp \left\{
  \frac{\tilde{\Phi}(\gamma')}{\kB T} \right\} d\gamma'
  \nonumber \\
  && + \frac{1}{w} \int\limits_{-1/2}^\gamma \exp \left\{
  \frac{\tilde{\Phi}(\gamma')}{\kB T} \right\} d\gamma' \Bigg)
  d\gamma \Bigg]^{-1} D \left( w - \frac{1}{w} \right) .
\label{plasticsm2}
\end{eqnarray}
For small $\varepsilon$, this result is consistent with
Eqs.~(\ref{rdefmod}) and (\ref{drift}). A Kramers-type formula
\citep{Kramers40} for the rate of transitions between reference
states thus follows immediately from the periodicity requirement.
In some analytical developments, but certainly also in numerical
calculations, it is more convenient to obtain a solution of the
diffusion equation on a compact interval with periodic boundary
conditions rather than a drifting and diffusively broadening
solution on the real axis.

\subsection{Thermodynamic formulation}
At this point we know that a complete and elegant kinetic theory
can be constructed on the quotient group of relative deformations
with respect to a slowly evolving reference state. However,
several further questions remain to be addressed: Can one find
thermodynamic building blocks directly for the quotient group
model? If so, how are these building blocks related to those of
the model on the cover group? Are there any extra rules or recipes
that need to be added to the thermodynamic framework?

An obvious additional recipe is the periodicity requirement for
$p_\bbbzi$. This periodicity cannot be derived directly from the
thermodynamic consistency on the cover group; rather, periodicity
is a consequence of superposing all smoothly patched solutions on
the different equivalent intervals. In that sense, the periodicity
requirement is a fundamentally new ingredient of the thermodynamic
approach resulting from the act of covering. For generalizing the
periodicity requirement to more complicated situations, it is
important to remember that there exists a well-developed theory of
harmonic analysis on locally compact groups (with particularly
powerful results for commutative or compact groups).

We now turn to the thermodynamic building blocks. The
conformational contribution to the energy in Eq.~(\ref{toyE}) per
unit channel width, $(1/a) \langle \! \langle \Phi \rangle \!
\rangle_{\rm sp}$, can be rewritten as $(1/a) \langle \! \langle
\Phi \rangle_\bbbzi \rangle_{\rm sp}$ because the potential energy
is invariant under shear transformations from the discrete group
$\bbbz$. The discussion of entropy can be based on the following
identity:
\begin{eqnarray}
  \int\limits_{-\infty}^\infty p(\gamma) \ln p(\gamma) d\gamma &=&
  \int_{I_0} p_\bbbzi(\gamma) \ln p_\bbbzi(\gamma) d\gamma
  + \sum_{z\in\bbbzi} p_z \ln p_z \nonumber\\
  &+& \sum_{z\in\bbbzi} p_z \int_{I_z} \frac{p(\gamma)}{p_z}
  \ln \frac{p(\gamma)}{p_z p_\bbbzi(\gamma-z)} d\gamma .
\label{entropident}
\end{eqnarray}
The first term on the right-hand side of Eq.~(\ref{entropident})
represents the entropy associated with the periodic probability
density $p_\bbbzi$ on the interval $I_0$, whereas the second term
results from the distribution over the equivalent intervals $I_z$.
The last term vanishes if the normalized probability on each
interval $I_z$, that is $p(\gamma)/p_z$, is equal to the properly
shifted distribution $p_\bbbzi$. Any deviation of $p(\gamma)/p_z$
from the shifted $p_\bbbzi$ leads to a reduction of entropy. This
loss of entropy cannot be estimated with the knowledge of
$p_\bbbzi$ and $p_z$ alone. Nevertheless, the dynamics on the
interval $I_0$ can be described rigorously. Therefore, the entropy
term resulting from a lack of proportionality of the distributions
on the different intervals $I_z$ can only affect the transitions
between the different intervals. The transition regimes between
different intervals are not self-similar and cannot be obtained
from the theory on the quotient group.

After discussing the conformational contributions to the energy
and entropy in terms of $p_\bbbzi$, we need to calculate their
functional derivatives. In particular, we need to analyze the
significance of the constraint (\ref{constraintstrain}) on the
average shear deformation and the reciprocal condition
(\ref{LAcons2}) in defining functional derivatives for the
quotient group model. The constraint condition (\ref{LAcons2}) was
recognized to be crucial to obtain the proper diffusion equation;
we hence need to take it over into the model on the quotient
group. In calculating $\delta E/\delta p$, this constraint leads
to a jump of $\delta E/\delta p$ rather than periodic boundary
conditions. Such a jump is actually crucial to obtain periodic
steady-state solutions $p_\bbbzi$ in the presence of a flux. Some
overall effect of the jumps in the probability levels $p_z$ of the
smoothly patched solutions on the different intervals $I_z$ due to
a flux in a driven system, which is lost in defining $p_\bbbzi$ by
superposition, is reintroduced through the functional derivatives
with jumps. Note that the functional derivatives are modified such
that there occurs a jump between $\gamma = - 1/2$ and $\gamma = +
1/2$, but the periodicity of all derivatives with respect to
$\gamma$ remains unaffected. The proper definition of functional
derivatives has to be seen together with the new periodicity
requirement. In the original direct construction of a kinetic
theory on the interval $I_0$ \citep{hco176}, the counterpart of
the constrained functional derivative is achieved through a formal
driving parameter which can be related to the macroscopic strain
rate and shear stress by imposing the constraint that the applied
strain rate is the sum of the elastic and plastic strain rates.

We finally need to discuss the Poisson and dissipative brackets.
The Poisson bracket results from a Lie algebra and hence coincides
for the group $I_0$ and its cover $\bbbr$. Also the dissipative
bracket in Eq.~(\ref{toydissB}) involves only local quantities and
an average of a periodic function, so that it may be used directly
on the quotient group. When all the thermodynamic building blocks
are used in the fundamental equation (\ref{LMformulation}), we
recover the equations of Sect.~\ref{secke}. Periodicity
supplemented by properly defined functional derivatives are the
new ingredients into the thermodynamic framework.

\section{Summary and discussion}\label{secdiscussion}
The present paper can be considered from two different
perspectives. From a mathematical perspective, it analyses the
freedom in associating Lie groups with the Lie algebras that
provide the structure required to introduce reversible time
evolution. From a thermodynamic perspective, this paper develops
the tools for models with multiple equivalent reference states
evolving slowly in time.

All continuous groups obtained as quotients of a universal
covering group and a discrete normal subgroup of the covering
group possess the same Lie algebra. The discrete normal subgroup,
which may be finite or infinite, may be interpreted as a set of
equivalent reference states, and the quotient group describes
deformation effects with respect to a reference state. For a toy
model of crystal viscoplasticity, we have shown in detail how
modeling on the cover and quotient groups works, and how they are
related. On quotient groups, the new ingredients to thermodynamic
modeling are suitable periodicity requirements supplemented by
properly defined functional derivatives. Important mathematical
ingredients in addition to the correspondence between Lie groups
and Lie algebras are Haar measures and harmonic analysis on
locally compact groups.

\begin{figure}
\centerline{\scalebox{0.5}{\includegraphics{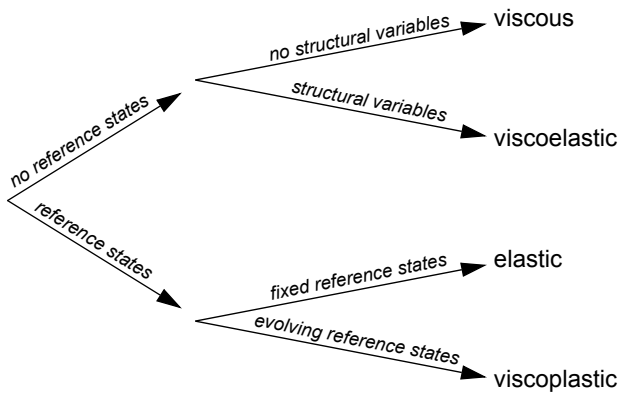}}}
\caption[ ]{Attributes of rheological models.}
\label{fig_elas_plas}
\end{figure}

The ideas of this paper naturally lead to a classification of
rheological models, depending on whether or not we make use of
reference states (see Fig.~\ref{fig_elas_plas}). From an abstract
point of view, reference states occur when the Lie group providing
the structure required to set up thermodynamics is not simply
connected so that it can be written as the quotient of the
universal covering group and a discrete group of reference states.
In particular, we obtain a clear distinction between
viscoplasticity and viscoelasticity, which often is not very
transparent in the literature, in terms of reference states. In
the absence of reference states we obtain the models of
viscoelasticity which have been applied successfully to many
complex fluids. The possibility of transitions between equivalent
reference states is the hallmark of viscoplasticity.

Viscoplasticity is often defined as rate-dependent plasticity,
where plasticity generally means the ability to change or deform
permanently (in contrast to elasticity, which refers to the
ability to change temporarily and revert back to the original
form). These permanent changes occur by transitions from reference
states to equivalent ones. Within the domain of a particular
reference state, deformations are reversible and hence elastic.
Only when a sufficiently high stress level is reached, transitions
between equivalent reference states occur with a noticeable rate
(in principle, transitions are possible even in the absence of
stresses, but they are extraordinarily rare). This stress level is
known as the yield point or, for anisotropic materials, as the
yield surface. As yield is a matter of ``noticeability,''
phenomenological definitions of viscoplasticity based on this
concept may not be ideal. The deformation behavior described by
long-living reference states is often referred to a
elastic-viscoplastic, whereas transitions at a noticeable rate
lead to a plastic flow associated with a very large viscosity.

The deformation with respect to evolving reference states enters
many models of viscoplasticity in which the total deformation is
split into a plastic and an elastic part, where only the elastic
part is used as a structural variable (see
\citet{Leonov76,HutterTervoort08} and references therein). Such a
splitting corresponds directly to the factorization of a covering
group. Usually, the relative or elastic deformation is treated by
explicit variables, whereas the transition rate between reference
states merely contributes to the rate of change of the relative
deformation variable. The relative deformation measures used in
previous phenomenological models of viscoplasticity correspond to
moments of the configurational distribution function in the
kinetic theory of the present paper, just as the conformation
tensors of the Maxwell and Oldroyd models of viscoelasticity
correspond to second moments of the connector vector of a Hookean
dumbbell model in polymer kinetic theory \citep{BirdetalDPLII}.
Even if a kinetic theory on the quotient group does not treat
reference states explicitly, the suggested analysis of the group
theoretical background clearly shows the underlying existence of
reference states and hence the viscoplastic nature of the models.
Group theory thus helps to identify viscoplastic models, and the
phenomena described by such models in a meaningful way are then
recognized as viscoplastic.

Kinetic theories of viscoplasticity, such as the toy model
employed for the purpose of illustration, are a promising starting
point for deriving existing phenomenological models and for
generalizing them, both for crystalline and for amorphous materials.
They offer a link between detailed atomistic
considerations \citep{ElAzab00,Arsenlisetal04,Johnsonetal07} and
phenomenological models \citep{Rice71,PanRice83,BoyceParksArg88}.
For deriving kinetic theories by statistical mechanics
\citep{hcobet,hco101,hco131,hco173}, it is important to realize
the occurrence of discrete sets of reference states and of
transition rates. The statistical mechanics of models with
reference states should hence be expected to be based on
Kramers-type formulas \citep{Kramers40} for transition rates.

\section*{Acknowledgment}
I gratefully acknowledge many illuminating discussions with Markus
H\"utter.


\end{document}